\documentclass
[secnumarabic,amsmath,amssymb,showpacs,nobibnotes,aps, pre, 12pt]{revtex4-1}

\usepackage{amssymb}

\usepackage{graphicx}
\usepackage{dcolumn}
\usepackage{bm}
\usepackage{color}

\begin{document}
\title{Spontaneous toroidal rotation, anomalous radial particle flux, and the electron-ion asymmetric anomalous viscous damping}
\author{Shaojie Wang}
\email{wangsj@ustc.edu.cn}
\affiliation{Department of Modern Physics, University of Science and Technology of China, Hefei, 230026, China}
\date{\today}

\begin{abstract}
A spontaneous toroidal rotation due to the electron-ion asymmetric anomalous viscous damping and the turbulent radial particle flux has been found, which explains the experimental observation of the anomalous toroidal momentum source in the edge of a tokamak plasma.
\end{abstract}


\maketitle

\section{INTRODUCTION}
Spontaneous spin-up or flow generation is of great interest in many fields of scientific research, such as the jet-stream formation in the atmosphere \cite{BaldwinScience07}, the giant zonal belt on the Jupiter \cite{PerezIca12}, the flow pattern generated in a turbulent neutral fluid \cite{FrischBook95}, the dynamo effect in a conducting fluid \cite{MoffattBOOK78}, the particle acceleration in solar flares \cite{PriestAAR02}, and the zonal flows in a tokamak magnetic fusion plasma \cite{LinScience98}. The toroidal rotation is crucially important for the tokamak fusion reactor, a magnetic fusion torus, such as the International Thermonuclear Experimental Reactor (ITER) \cite{DoyleNF07}, since it can stabilize the macroscopic instabilities \cite{BondesonPRL94,BettiPRL95}, and regulate the microturbulence \cite{BiglaryPoFB90,ConnorPPCF94,TerryRMP00,WangPRL06a}. Recent tokamak experiments suggest that an intrinsic toroidal rotation exists \cite{PeetersNF11, AngioniNF12,LeePRL03,RiceNF07,RicePPCF08,SolomonPPCF07,SolomonNF09}, due to an anomalous momentum source in the edge region, which can not be explained by the effects of momentum pinch \cite{PeetersPRL07,HahmPoP07}; the effects of the residual Reynolds stress due to parallel symmetry-breaking on the parallel momentum generation \cite{DiamondPoP08} and the effects of toroidal symmetry-breaking on the toroidal acceleration \cite{GarbetPoP13} have been discussed.

In this paper, we shall show that the intrinsic toroidal rotation can be generated by the turbulent (anomalous) radial particle flux, which is reminiscent of the bootstrap current generated by the neoclassical particle radial diffusion \cite{GaleevJETP68,BickertonNPS71}; the proposed mechanism of toroidal momentum generation can be explained by the electron-ion asymmetric anomalous viscous damping, which does not need the spatial symmetry-breaking, such as the toroidal symmetry-breaking or the parallel symmetry-breaking. This paper is organized as follows. In Sec. II, we review the fundamental theory and the related experiments; in Sec. III, we present the theoretical analysis; in Sec. IV, we discuss the applications of the proposed model; in Sec. V, we present the summary and discussions.

\section{Review of the fundamentals}

\subsection{Fundamental theory}

The equilibrium magnetic field of a tokamak is written as $\bm B =I\left(r\right)\nabla \zeta +\nabla \zeta \times \nabla \psi \left(r\right)$, with $I=R B_T$, $B_T$ the toroidal magnetic field, $R$ the major radius. The poloidal magnetic flux is $\psi\left(r\right)$, with $r$ essentially the minor radius of the torus. In the large-aspect-ratio limit ($r\ll R$), the poloidal magnetic field is given by $B_{P}=\psi'/R$, with the prime denoting the derivative with respect to $r$. $\zeta$ is the toroidal angle.

The two-fluid mean-field momentum equation for a simple electron-deuteron plasma can be written as \cite{HazeltineBOOK92},
\begin{equation}
 \partial_t\left(n_s m_s \bm {u}_s\right) +\nabla\cdot\left(n_s m_s \bm {u}_s \bm {u}_s + p_s\bm {I}+\bm{\Pi}_{s,n} +\bm{\Pi}_{s,a} \right) =n_s e_s \left[-\nabla\Phi\left(\psi\right)+ E_{\zeta}\nabla \zeta +\bm {u}_s \times \bm B\right]+\bm {F}_s ,\label{eq:momentum}
\end{equation}
where the subscript $s=e,i$ denotes the charged particle species. $m$ and $e$ are particle mass and charge, respectively. $n$ and $\bm u$ are density and fluid velocity, respectively. Note that the quasi-neutrality condition demands that $n_e=n_i=n$. $-\nabla\Phi\left(\psi\right)$ and $E_{\zeta}\nabla \zeta=\bm E_T$ are the radial electric field and the toroidal inductive electric field, respectively. $p$ is the pressure. $\bm {\Pi}_{n}$ is the CGL form of the stress tensor, which includes the neoclassical effects \cite{HazeltineBOOK92}. $\bm{\Pi}_a$ is the anomalous viscosity tensor, which accounts for the anomalous transport of momentum due to electrostatic turbulence, therefore, it includes the usual Reynolds stress term \cite{CallenNF09,GarbetPoP13}. $\bm {F}$ is the collisional friction force. In Eq. (\ref{eq:momentum}), we have neglected the source term due to the external momentum injection. Note that all quantities in Eq. (\ref{eq:momentum}) are turbulence-ensemble averaged.

Acting $R^2\nabla\zeta\cdot$ on Eq. (\ref{eq:momentum}) and averaging the resulting equation over the magnetic flux surface of a tokamak yields the toroidal angular momentum equation \cite{HazeltineBOOK92}
\begin{equation}
 \partial_t\left\langle n_s m_s u_{s,\zeta}\right\rangle +\frac{1}{\mathcal{V}'}\frac{d}{dr}\left(\mathcal{V}' \left\langle n_s m_s u_s^r u_{s,\zeta}+\Pi_{s,\zeta}^r \right\rangle\right)-\left(\left\langle F_{s,\zeta}\right\rangle+n_s e_s E_{\zeta}\right)= e_s \Gamma_s^r \frac{d \psi}{dr},\label{eq:tam}
\end{equation}
where $\mathcal{V}\left(r\right)$ is the volume enclosed in the magnetic flux surface labeled by $r$.  Note that the CGL form of the neoclassical viscosity tensor does not survive the flux surface average \cite{HazeltineBOOK92}. The radial particle flux is $\Gamma ^r=n\left(r\right)\left\langle u^r\right\rangle$. Note that $u^r=\nabla r\cdot \bm u$, $u_{\zeta}=R^2\nabla \zeta \cdot \bm u$, and $\Pi_{\zeta}^{r}=\nabla r\cdot\bm{\pi}\cdot R^2\nabla\zeta$.

A simple interpretation of Eq. (\ref{eq:tam}) is as follows. The radial particle flux can be decomposed into two components: the neoclassical radial particle flux $\Gamma_n$ and the anomalous radial particle flux $\Gamma_a$.
\begin{equation}
\Gamma_s^r=\Gamma^r_{s,n}+\Gamma^r_{s,a}.\label{eq:gamma-r}
\end{equation}

The neoclassical radial particle flux is driven by the toroidal collisional friction force and the toroidal inductive electric field,
\begin{equation}
e_s\Gamma_{s,n}^r\psi'=-\left(\left\langle F_{s,\zeta}\right\rangle+n_s e_s E_{\zeta}\right);\label{eq:pneo}
\end{equation}
the anomalous radial particle flux is driven by the anomalous toroidal force written in terms of the anomalous toroidal viscosity and the inertial term,
\begin{equation}
e_s\Gamma_{s,a}^r\psi'=\partial_t\left\langle n_s m_s u_{s,\zeta}\right\rangle +\frac{1}{\mathcal{V}'}\frac{d}{dr}\left(\mathcal{V}' \left\langle n_s m_s u_s^r u_{s,\zeta}+\Pi_{s,\zeta}^r \right\rangle\right). \label{eq:pano}
\end{equation}

This interpretation of the anomalous radial particle flux is consistent with the previous theories (Sec. 4 of Ref. \onlinecite{CallenNF09} and Sec. II. C of Ref. \onlinecite{GarbetPoP13}); the toroidal momentum equation used in Ref. \onlinecite{GarbetPoP13} includes an additional term accounting for the toroidal turbulence acceleration, which depends on the toroidal symmetry breaking and shall be ignored in this paper. Following Ref. \onlinecite{CallenNF09, GarbetPoP13}, one concludes that although Eq. (\ref{eq:tam}) was originally derived for an quiescent plasma \cite{HazeltineBOOK92}, it is also applicable for a turbulent plasma with the effects of Reynolds stress included in the anomalous viscosity term.

It is useful to make some comments on the neoclassical bootstrap current in the large-aspect-ratio limit. The collisional friction force is given by \cite{HirshmanNF81}
\begin{equation}
\bm {F}_i=\bm{F}_{ie}=-\bm{F}_{ei}=-\nu_{ei} n m_e\left(\bm{u}_i-\bm{u}_e\right),\label{eq:friction}
\end{equation}
with $\nu_{ei}$ the electron-ion collision frequency.

The neoclassical radial particle radial flux, which satisfies the auto-ambipolarity condition, is given by \cite{HazeltineBOOK92}
\begin{equation}
 \Gamma_{i,n}^r=\Gamma_{e,n}^r\sim-f_t\nu_{ei}\rho_{eP}^2\frac{dn}{dr}-f_t n E_{\zeta}\psi'/R^2,\label{eq:gamman}
\end{equation}
with $f_t$ the trapping ratio. $\rho_{eP}=\sqrt{2m_e T}/e_e B_P$ is the electron poloidal gyro-radius, with $T$ the temperature of the plasma.

Substituting Eq. (\ref{eq:friction}), Eq. (\ref{eq:gamman}) into Eq. (\ref{eq:pneo}), one finds
\begin{equation}
\left\langle n_e e_i \left(\bm{u}_i-\bm{u}_e\right)\right\rangle_{\zeta}\frac{1}{R} \sim -f_t \frac{T}{B_P}\frac{dn}{dr}+\left(1-f_t\right)\sigma_S E_{T},\label{eq:bs}
\end{equation}
with $\sigma_s$ the Spitzer conductivity. This shows the well-known bootstrap current and the neoclassical reduction of the Spitzer conductivity.

\subsection{Experiments related to the friction term}

Both of the toroidal rotation and the toroidal current are crucially important for a tokamak fusion reactor. Fortunately, the toroidal current can be partly provided by the bootstrap current driven by the radial neoclassical particle flux. The neoclassical bootstrap current and electric conductivity depends on the friction term, which has been discussed in the above. The friction term is important in the toroidal momentum equation. Therefore, it is useful to examine the experiments related to the friction term.

Although generally tokamak experiments indicate that the radial particle flux is anomalous (typically 30 times larger than the friction-generated neoclassical radial particle flux), one can attribute the anomaly of the radial particle flux to an anomalous toroidal force (due to anomalous toroidal viscosity). However, it should be pointed out that many experiments have been carried out to test the neoclassical theory of bootstrap current and the Spitzer conductivity \cite{ZarnstorffPRL88,HwangPRL96,ThomasPRL04,WadePRL04}. These experimental observations indicate that the toroidal current is related to the toroidal inductive electric field and the radial pressure gradient in the way predicted by the neoclassical theory based on the collisional friction. Although there are theoretical arguments that the bootstrap current may be enhanced by the effect of radial electric field \cite{KaganPRL10}, or by the effect of turbulence scattering \cite{McDevittPRL13}, the order of magnitude of the bootstrap current and hence the friction does not change. Therefore, one concludes from the experimental observations that the toroidal friction term is not anomalous; the toroidal friction force acted on one species is balanced by the poloidal magnetic field crossed with the radial current carried by its neoclassical flux. Eq. (\ref{eq:bs}) has been validated by experiments.

\subsection{Experiments related to the intrinsic rotation}

Intrinsic toroidal rotation has been observed in many tokamak experiments \cite{PeetersNF11, AngioniNF12,LeePRL03,RiceNF07,RicePPCF08,SolomonPPCF07,SolomonNF09}, which indicate that significant co-current toroidal rotation exists in tokamaks without any external momentum injection; this important phenomena is difficult to understand, especially when considering that a tokamak is essentially a toroidal symmetric system.

In C-mod, during a transient phase of density ramp-up, a toroidal counter-current acceleration is observed for both the ion-cyclotron-resonance heated plasma and the lower-hybrid wave-driven plasma, and a co-current acceleration is observed during a density ramp-down phase \cite{RicePPCF08}. The most convincing evidence of the anomalous torque and hence the spontaneous rotation may be the recent measurement of the anomalous co-current torque reported by the DIII-D team \cite{SolomonPPCF07,SolomonNF09}; the anomalous torque density is in the co-current direction, with its peak value $\sim0.7Nm/m^3$ located at the edge.

The observation of an anomalous toroidal torque in the edge of DIII-D \cite{SolomonPPCF07,SolomonNF09} is consistent with the observation of spontaneous toroidal rotation in C-mode \cite{LeePRL03}, where a model equation is used to interpret the toroidal rotation.

\begin{subequations}
\begin{eqnarray}
&&\partial_t P +\frac{1}{r}\partial_r\left[r\left(-\chi_{\zeta}\partial_r P -\frac{r}{a}v_c P\right)\right]=0,\label{eq:Lee1}\\
&&P\left(r=a\right)=P_0,\label{eq:Lee2}
\end{eqnarray}
\end{subequations}
with $P$ the toroidal momentum of the plasma, $a$ the minor radius of the device, and $v_c$ the momentum pinch velocity. Note that a finite toroidal momentum at the edge in Eq. ({\ref{eq:Lee2}}) implies a source at the edge when a no-slip boundary condition is applied.

\section{Theoretical analysis}
Consider a large-aspect-ratio tokamak. Eq. (\ref{eq:tam}) is reduced to
\begin{equation}
\partial_t \left( n_s m_s U_s \right)+\frac{1}{r}\frac{d}{dr} \left\{ r n_s m_s\left[ \left( u_s^r + u^r_{s,mp} \right)  U_s- \chi_{s,\zeta}\frac{d}{dr} U_s \right] \right\}-R\left(F_{s,\zeta}+n_se_sE_{\zeta}\right)= e_s\Gamma_s^r B_P,\label{eq:tm}
\end{equation}
where $U_s=R\nabla\zeta\cdot\bm{u}_s$ is the toroidal velocity. $\chi_{s,\zeta}$ is the momentum diffusivity. $u^r_{s,mp}$ is the radial pinch velocity of the toroidal momentum.  Note that the anomalous toroidal viscosity is written in a momentum diffusion term and a momentum pinch term.

The ion momentum pinch is given by \cite{PeetersPRL07,HahmPoP07}
\begin{equation}
u^r_{_i,mp} = -\alpha \frac{1}{R}\chi_{i,\zeta}, \label{eq:mp}
\end{equation}
with $\alpha=2, 4$, to take into account of the momentum pinch due to the Coriolis force \cite{PeetersPRL07,HahmPoP07}. For simplicity, we shall assume that this toroidal momentum pinch-diffusion relation is also applied to electrons.

The neoclassical particle flux and the anomalous particle flux are written as
\begin{equation}
\Gamma_{s,n}^r=-\frac{1}{e_sB_P}R\left(F_{s,\zeta}+n_se_sE_{\zeta}\right),\label{eq:gamma-n}
\end{equation}
\begin{equation}
\Gamma_{s,a}^r=\frac{1}{e_sB_P}\partial_t \left( n_s m_s U_s \right)+\frac{1}{e_sB_P}\frac{1}{r}\frac{d}{dr} \left\{ r n_s m_s\left[ \left( u_s^r -\frac{\alpha}{R}\chi_{s,\zeta} \right)  U_s- \chi_{s,\zeta}\frac{d}{dr} U_s \right] \right\}.\label{eq:gamma-a}
\end{equation}

The ion radial particle transport is schematically shown in Fig. 1.


The ambipolarity condition, which is derived from Ampere's law, demands that
\begin{equation}
e_e\Gamma^r_e+e_i\Gamma^r_i=0.\label{eq:amb}
\end{equation}
Since the neoclassical particle flux is auto-ambipolar, the ambipolarity condition is reduced to
\begin{equation}
\Gamma^r_{e,a}=\Gamma^r_{i,a}.\label{eq:amb-r}
\end{equation}

For typical tokamak experimental parameters, $U_i\sim 3\times10^4m/s$, $U_e\sim 10^6m/s$,
\begin{equation}
\frac{m_i U_i}{m_e U_e}\sim 10^2. \label{eq:V}
\end{equation}
Following Eq. (\ref{eq:gamma-a}) and Eq. (\ref{eq:amb-r}), one concludes that
\begin{equation}
\frac{\chi_{e,\zeta}}{\chi_{i,\zeta}}\sim 10^2, \label{eq:chi-e}
\end{equation}
which indicates an electron-ion asymmetry of anomalous toroidal viscous damping.
\subsection{Relaxation process}

For typical tokamak experimental parameters, $\nu_{ei}\sim10^5/s$; the ion toroidal momentum confinement time is $\tau_{i,\zeta}\sim a^2/\chi_{i,\zeta}\sim 10^{-1}s$; $\Gamma^r/na=u^r/a\sim 1m/s$. Following Eq. (\ref{eq:chi-e}), one finds $\tau_{e,\zeta}\sim a^2/\chi_{e,\zeta}\sim 10^{-3}s$. Using these parameters, one finds that the electron momentum convection term due to the particle flux can be safely dropped.

We write down the electron toroidal momentum equation
\begin{equation}
\partial_t \left( n_e m_e U_e \right)
+\frac{1}{r}\frac{d}{dr} \left\{ r n_e m_e \chi_{e,\zeta}\left[-\frac{\alpha}{R} - \frac{d}{dr} \right] U_e \right\}
-n_e\left[ \nu_{ei}m_e \left(U_e-U_i\right)+ e_e E_T\right]
= e_e \left(\Gamma^r_{e,n}+\Gamma^r_{e,a}\right) B_P.\label{eq:tme}
\end{equation}

We observe that
\begin{equation}
1/\nu_{ei}\ll \tau_{e,\zeta}\ll\tau_{i,\zeta},
\end{equation}
which indicates that the collisional relaxation of $U_e$ is faster than the turbulence relaxation.

On the short time scale of $1/\tau_{ei}$, $U_e$ is quickly relaxed through the neoclassical collisional process,
\begin{equation}
\nu_{ei}n_em_e\left(U_e-U_i\right)+n_e e_e E_T=e_e\Gamma^r_{e,n}B_P,\label{eq:gamma-en}
\end{equation}
which is consistent with Eq. (\ref{eq:bs}) and Eq. (\ref{eq:gamma-n}).

On the longer time scale of $\tau_{e,\zeta}$, $U_e$ is relaxed through the turbulent process to generate the anomalous radial electron flux
\begin{equation}
\frac{1}{r}\frac{d}{dr} \left\{ r n_e m_e\left[ -\frac{\alpha}{R}\chi_{e,\zeta} U_e  - \chi_{e,\zeta}\frac{d}{dr} U_e \right] \right\}
= e_e \Gamma^r_{e,a}B_P,\label{eq:gamma-ea}
\end{equation}

On the time scale of $\tau_{i,\zeta}$, the ion toroidal momentum is relaxed to generate the anomalous radial ion flux
\begin{equation}
\partial_t \left(n_i m_i U_i\right)+\frac{1}{r}\frac{d}{dr} \left\{ r n_i m_i\left[ -\frac{\alpha}{R}\chi_{i,\zeta} U_i  - \chi_{i,\zeta}\frac{d}{dr} U_i \right] \right\}
= e_i \Gamma^r_{i,a}B_P.\label{eq:gamma-ia}
\end{equation}

\subsection{Single-fluid toroidal momentum equation}
The single-fluid toroidal velocity $V$ is defined by $\rho V\equiv n M V \equiv n(m_i+m_e)V=n(m_iU_i+m_eU_e)$. The single-fluid toroidal momentum equation can be found by adding the electron-version of Eq. (\ref{eq:tm}) to its ion-version and using the ambipolarity condition. Since Eq. (\ref{eq:tm}) can be written as Eq. (\ref{eq:gamma-a}) and Eq. (\ref{eq:gamma-n}), and the ambipolarity condition is reduced to Eq. (\ref{eq:amb-r}), one can write down the single-fluid equation through combining Eq. (\ref{eq:amb-r}), Eq. (\ref{eq:gamma-a}) and Eq. (\ref{eq:gamma-n}),
\begin{equation}
\partial_t \left( \rho V \right)+\frac{1}{r}\frac{d}{dr} \left\{ r \rho \left[ \left( u^r -\frac{\alpha}{R}\chi_{i,\zeta} \right) - \chi_{i,\zeta}\frac{d}{dr} \right] V \right\}= -\frac{1}{r}\frac{d}{dr} \left[ r n m_e \left(\chi_{e,\zeta}-\chi_{i,\zeta}\right)\left( -\frac{\alpha}{R} - \frac{d}{dr}  \right)U_e  \right], \label{eq:tm-0}
\end{equation}
which is the exact result of the ambipolarity condition.

A single-fluid toroidal momentum equation consistent with Eq. (\ref{eq:Lee1}), which has been widely used in previous works \cite{LeePRL03,SolomonPPCF07,SolomonNF09} on the transport of toroidal momentum, can be found from Eq. (\ref{eq:tm-0}) by simply dropping its right-hand side. If $\chi_{e,\zeta}=\chi_{e,\zeta}$, this would have been true. If $\chi_{e,\zeta}\approx \chi_{e,\zeta}$, Eq. (\ref{eq:Lee1}) would have been a fairly good approximation for typical experiments for which $m_iU_i\gg m_e U_e$. However, according to the above discussions, for typical experiments, there this a strong electron-ion asymmetry of toroidal viscus damping [see, Eq. (\ref{eq:chi-e})].

Using Eq. (\ref{eq:chi-e}) and Eq. (\ref{eq:gamma-ea}), one finds that Eq. (\ref{eq:tm-0}) can be written as
\begin{equation}
\partial_t \left( \rho V \right)+\frac{1}{r}\frac{d}{dr} \left\{ r \left[ \left( u^r -\frac{\alpha}{R}\chi_{i,\zeta} \right) \rho V-\rho \chi_{i,\zeta}\frac{d}{dr} V \right] \right\}\approx -e_e\Gamma^r_{e,a} B_p . \label{eq:tm-1}
\end{equation}

Using Eq. (\ref{eq:amb-r}), and defining the anomaly factor
\begin{equation}
\Gamma^r_a=f_a \Gamma^r,
\end{equation}
one finds the single-fluid toroidal momentum equation
\begin{equation}
\partial_t \left( \rho V \right)+\frac{1}{r}\frac{d}{dr} \left\{ r \left[ \left( u^r -\frac{\alpha}{R}\chi_{i,\zeta} \right) \rho V-\rho \chi_{i,\zeta}\frac{d}{dr} V \right] \right\}= f_a e_i B_p \Gamma^r. \label{eq:tm-f}
\end{equation}
Note that typical experimental parameters indicate that $f_a\leq 1$.

General remarks should be made on Eq. (\ref{eq:tm-f}). The right-hand side identifies the anomalous momentum source observed in experiments. Note that usually the particle source is in the edge of a tokamak due to the shallow penetration of the fueling; therefore, in a steady-state, the particle flux is non-zero only in the edge region. This
explains why the experimental observations indicate that usually an anomalous toroidal momentum source is located in the edge region \cite{LeePRL03,RiceNF07}. The source term, the right-hand side of Eq. (\ref{eq:tm-1}) represents the intrinsic toroidal rotation driven by the anomalous radial particle flux, which is different from the previous models which need a spatial (essentially toroidal) symmetry-breaking effect\cite{DiamondPoP08,GarbetPoP13}.

The physical mechanism of spontaneous momentum generation behind Eq. (\ref{eq:tm-f}) can be understood by examining Eq. (\ref{eq:tm}). The friction forces acted on electrons and ions exactly cancel out due to momentum conservation, and the forces due to the toroidal inductive electric field cancel out through the quasi-neutrality condition. Although the toroidal component of magnetic Lorentz forces ($e_s \Gamma^r_s B_P$) cancel out due to the ambipolarity condition, the electron toroidal viscous damping rate ($\chi_{e,\zeta}$) is larger than the ion toroidal viscous damping rate ($\chi_{i,\zeta}$) [see, Eq. (\ref{eq:chi-e})], this electron-ion asymmetry of viscous damping can generate a net momentum for the two-component system.

The spontaneous ion toroidal rotation can also be understood in a two-fluid picture. From the ion-version of Eq. (\ref{eq:gamma-a}), one can see that the anomalous toroidal ion viscous force is balanced by the product of the anomalous radial ion current and the poloidal magnetic field, which is schematically shown in Fig. 1.

\section{APPLICATIONS}
If one assumes a constant value of $\chi_{i, \zeta}$, the analytic solution of Eq. (\ref{eq:tm-f}) with a no-slip boundary condition can be readily found, and the toroidal velocity is given by
\begin{equation}
 U\left( r \right)= \left[ c \left( a \right)- c \left( r \right) \right] \exp \left(\int_0^r dr\frac{u^r+u^r_{i,mp}}{\chi_{i,\zeta}}\right) , \label{eq:solution}
\end{equation}
where $a$ is the minor radius of the boundary,
\begin{equation}
 c\left( r \right)=\int_0^r dr \frac{S_{\zeta}}{r n m_i \chi_{i,\zeta}} \exp \left(-\int_0^r dr\frac{u^r+u^r_{i,mp}}{\chi_{i,\zeta}}\right) , \label{eq:c}
\end{equation}
\begin{equation}
 S_{\zeta} \left( r \right)= \int_0^r rdr n e_i u^r B_P. \label{eq:sm}
\end{equation}
$u^r$ can be determined as follows.

Particle balance demands
\begin{equation}
2\pi r n u^r=\int_0^r 2\pi rdr S_0g \left(r\right),
\end{equation}
with $S_0g\left(r\right)$ the particle source and $S_0$ a constant. $S_0$ can be determined by
\begin{equation}
\left[n u^r\right]_a 2\pi a =\pi a^2 \bar {n}/\tau_p,
\end{equation}
with $\bar {n}$ the volume averaged density.

Typical parameters of an ITER-like tokamak are as follows. $R/a=6.2m/2m$, $B_T=5.3T$, $e_i=1.6\times10^{-19}C$, $m_i=2.5m_p$. The energy confinement time is $\tau_E\approx 3.5 s$, the particle confinement time is $\tau_p\approx 10s$, which is roughly 3 times of the energy confinement time. We shall assume $n=n_0\left[1-\left(\frac{r}{a}\right)^2\right]^{0.5}+0.1n_0$, with $n_0=1.5\times10^{20}/m^3$, and $q\left(r\right)=\frac{rB_T}{RB_P}=1.1\left[1+2.5\left(\frac{r}{a}\right)^2\right]$. Since the toroidal momentum confinement time is usually similar to the energy confinement time \cite{LeePRL03}, we shall assume $\chi_{i,\zeta}\approx\frac{a^2}{5.8\tau_E}$. A Gaussian type of particle source profile is assumed, which is non-zero in the edge region $0.9<r/a<1$.

The particle source $S_0g\left(r\right)$ and the toroidal force density $n e_i u^r B_P$ are shown in Fig. 2.


With the above parameters, the toroidal rotation speed for different momentum diffusivities ($\chi_{\zeta}$) with different momentum pinch factors ($\alpha$) are shown in Fig. 3.


Note that the solution of the core rotation ($r<0.9$) shown in Fig. 3 can also be obtained by solving the source-free version of Eq. (\ref{eq:tm-f}) and adjusting the prescribed boundary condition $V\left(r=0.9\right)=U_0$, which is similar to the assumption used in Ref. \onlinecite{LeePRL03}.

Fig. 3 clearly demonstrates that a significant spontaneous toroidal rotation can be driven by the anomalous particle diffusion in an ITER-like tokamak fusion reactor, which is typically from $200km/s$ in the edge to $500km/s$ in the core, and this bootstrap rotation may be sufficient to stabilize the resistive wall mode \cite{KikuchiRMP12,RiceNF07}; it is interesting to note that the spontaneous co-current toroidal rotation predicted by this model roughly agrees with the value extrapolated from the present experimental data \cite{RiceNF07}. The effect of momentum pinch \cite{PeetersPRL07,HahmPoP07} can be seen by contrasting Fig. 3(a, b) to Fig. 3(c).

It is of interest to compare the proposed model with the recent experimental observations of the spontaneous rotation. In a steady-state plasma fueled at the edge, the particle flux is radially outward, thus produces a co-current torque at the edge; this prediction qualitatively agrees with the experimental observations of the spontaneous rotation of the H-mode plasma \cite{LeePRL03,RiceNF07,SolomonPPCF07}. The direction of the spontaneous rotation of the L-mode plasma is complicated \cite{RiceNF07}, and the counter-current spontaneous rotation of the L-mode plasma may be due to the neoclassical effects \cite{SolomonNF09, WangPPCF12}. During a transient phase of density ramping-up, a radially inward particle flux is expected, which should generate a counter-current torque, and this prediction qualitatively agrees with the experimental observations of the counter-current acceleration during the density ramping-up phase of both the ion-cyclotron-resonance heated plasma and the lower-hybrid wave-driven plasma \cite{RicePPCF08}. The most convincing evidence of the anomalous torque and hence the spontaneous rotation may be the recent measurement of the anomalous co-current torque reported by the DIII-D team \cite{SolomonPPCF07,SolomonNF09}; the anomalous torque density is in the co-current direction, with its peak value $\sim0.7Nm/m^3$ located at the edge. The main parameters of the DIII-D tokamak is $R/a=1.6m/0.6m$. For the experiments discussed in Ref. \onlinecite{SolomonPPCF07}, one may assume that the volume-averaged electron density is  $\sim 3\times10^{19}/m^3$, the particle confinement time is $\tau_p\sim 0.3s$, and the poloidal magnetic field at the edge is $\sim 0.15T$. With these parameters, following the above analysis, one finds that the peak value of the torque density at the edge is $\sim 1Nm/m^3$, which agrees well with the experimental value \cite{SolomonPPCF07,SolomonNF09}.

\section{SUMMARY AND DISCUSSIONS}
In summary, we have proposed a theoretical model of spontaneous toroidal rotation driven by the turbulent radial particle flux through the electron-ion asymmetric viscous damping. The proposed model explains the anomalous co-current momentum source observed in the tokamak edge, and predicts a significant toroidal rotation in ITER, which may mitigate the needs of momentum injection and modify the design of a tokamak fusion reactor. It is interesting to note that both the toroidal current and the toroidal rotation are crucially important for the steady-state operation of tokamak fusion reactor \cite{KikuchiRMP12}; the bootstrap current is driven by the neoclassical particle diffusion and the intrinsic toroidal ion rotation is driven by the anomalous radial particle flux.

The mechanism of spontaneous momentum generation in a two-component system with the asymmetric damping can be simply understood as follows. Consider a two-component system given by
\begin{subequations}
\begin{eqnarray}
d_t P_1=-\nu_1P_1-\nu_0\left(P_1-P_2\right)+F,\\
d_t P_2=-\nu_2P_2-\nu_0\left(P_2-P_1\right)-F,
\end{eqnarray}
\end{subequations}
where $P_s$ is the momentum of each component; $\nu_s$ is the viscous damping rate; $\nu_0$ is the inter-component friction rate; $F$ is the driving force acted on component 1, which is opposite to the driving of component 2. The initial values are set as $P_1=0=P_2$. The final steady-state total momentum of the system is given by
\begin{equation}
P_1+P_2=\frac{\nu_2-\nu_1}{\nu_1\nu_2+\left(\nu_1+\nu_2\right)\nu_0}F,
\end{equation}
which clearly demonstrates a net momentum generated by the asymmetry of viscous damping, when the driving forces cancel out.

The physical mechanism of the electron-ion asymmetry of toroidal viscous damping is still an open issue, although it appears to be consistent with experimental observations. This paper proposes a new way to look into the physics of spontaneous toroidal spin-up; the present researches concentrate on the ion viscosity and the effects of possible parallel or toroidal symmetry breaking on the ion dynamics, this paper suggests that the electron viscous damping is important. A possible explanation of the electron-ion asymmetric toroidal viscous damping is as follows. Consider the trapped electron mode. Since the mode is resonant with electrons instead of ions, there should be a resonant electron momentum diffusion and no resonant diffusion of the ion momentum; this shall be left for the future researches.

Figure Captions:

Fig. 1. Ion radial flux driven by anomalous toroidal viscous force and the neoclassical toroidal friction.

Fig. 2. Particle source and toroidal force density.

Fig. 3. Toroidal rotation speed for different momentum diffusivities. $\chi_0=a^2/5.8\tau_E$. (a) $\alpha=4$; (b) $\alpha=2$; (c) $\alpha=0$.

\begin{acknowledgments}
This work was supported by the National Natural Science Foundation of China under Grant No. 11175178, No. 11375196 and the National ITER program of China under Contract No. 2014GB113000.
\end{acknowledgments}

\nocite{*}


\begin{thebibliography}{37}%
\makeatletter
\providecommand \@ifxundefined [1]{%
 \@ifx{#1\undefined}
}%
\providecommand \@ifnum [1]{%
 \ifnum #1\expandafter \@firstoftwo
 \else \expandafter \@secondoftwo
 \fi
}%
\providecommand \@ifx [1]{%
 \ifx #1\expandafter \@firstoftwo
 \else \expandafter \@secondoftwo
 \fi
}%
\providecommand \natexlab [1]{#1}%
\providecommand \enquote  [1]{``#1''}%
\providecommand \bibnamefont  [1]{#1}%
\providecommand \bibfnamefont [1]{#1}%
\providecommand \citenamefont [1]{#1}%
\providecommand \href@noop [0]{\@secondoftwo}%
\providecommand \href [0]{\begingroup \@sanitize@url \@href}%
\providecommand \@href[1]{\@@startlink{#1}\@@href}%
\providecommand \@@href[1]{\endgroup#1\@@endlink}%
\providecommand \@sanitize@url [0]{\catcode `\\12\catcode `\$12\catcode
  `\&12\catcode `\#12\catcode `\^12\catcode `\_12\catcode `\%12\relax}%
\providecommand \@@startlink[1]{}%
\providecommand \@@endlink[0]{}%
\providecommand \url  [0]{\begingroup\@sanitize@url \@url }%
\providecommand \@url [1]{\endgroup\@href {#1}{\urlprefix }}%
\providecommand \urlprefix  [0]{URL }%
\providecommand \Eprint [0]{\href }%
\providecommand \doibase [0]{http://dx.doi.org/}%
\providecommand \selectlanguage [0]{\@gobble}%
\providecommand \bibinfo  [0]{\@secondoftwo}%
\providecommand \bibfield  [0]{\@secondoftwo}%
\providecommand \translation [1]{[#1]}%
\providecommand \BibitemOpen [0]{}%
\providecommand \bibitemStop [0]{}%
\providecommand \bibitemNoStop [0]{.\EOS\space}%
\providecommand \EOS [0]{\spacefactor3000\relax}%
\providecommand \BibitemShut  [1]{\csname bibitem#1\endcsname}%
\let\auto@bib@innerbib\@empty
\bibitem [{\citenamefont {Baldwin}\ \emph {et~al.}(2007)\citenamefont
  {Baldwin}, \citenamefont {Rhines}, \citenamefont {Huang},\ and\ \citenamefont
  {McIntyre}}]{BaldwinScience07}%
  \BibitemOpen
  \bibfield  {author} {\bibinfo {author} {\bibfnamefont {M.~P.}\ \bibnamefont
  {Baldwin}}, \bibinfo {author} {\bibfnamefont {P.~B.}\ \bibnamefont {Rhines}},
  \bibinfo {author} {\bibfnamefont {H.~P.}\ \bibnamefont {Huang}}, \ and\
  \bibinfo {author} {\bibfnamefont {M.~E.}\ \bibnamefont {McIntyre}},\
  }\href@noop {} {\bibfield  {journal} {\bibinfo  {journal} {Science}\ }\textbf
  {\bibinfo {volume} {315}},\ \bibinfo {pages} {467} (\bibinfo {year}
  {2007})}\BibitemShut {NoStop}%
\bibitem [{\citenamefont {Perez-Hoyos}\ \emph {et~al.}(2012)\citenamefont
  {Perez-Hoyos}, \citenamefont {Sanz-Requena}, \citenamefont
  {Barrado-Izagirre}, \citenamefont {Rojas}, \citenamefont {Sanchez-Lavega},\
  and\ \citenamefont {Team}}]{PerezIca12}%
  \BibitemOpen
  \bibfield  {author} {\bibinfo {author} {\bibfnamefont {S.}~\bibnamefont
  {Perez-Hoyos}}, \bibinfo {author} {\bibfnamefont {J.~F.}\ \bibnamefont
  {Sanz-Requena}}, \bibinfo {author} {\bibfnamefont {N.}~\bibnamefont
  {Barrado-Izagirre}}, \bibinfo {author} {\bibfnamefont {J.~F.}\ \bibnamefont
  {Rojas}}, \bibinfo {author} {\bibfnamefont {A.}~\bibnamefont
  {Sanchez-Lavega}}, \ and\ \bibinfo {author} {\bibfnamefont {T.~I.}\
  \bibnamefont {Team}},\ }\href@noop {} {\bibfield  {journal} {\bibinfo
  {journal} {Icarus}\ }\textbf {\bibinfo {volume} {217}},\ \bibinfo {pages}
  {256} (\bibinfo {year} {2012})}\BibitemShut {NoStop}%
\bibitem [{\citenamefont {Frisch}(1995)}]{FrischBook95}%
  \BibitemOpen
  \bibfield  {author} {\bibinfo {author} {\bibfnamefont {U.}~\bibnamefont
  {Frisch}},\ }\href@noop {} {\emph {\bibinfo {title} {Turbulence}}}\ (\bibinfo
   {publisher} {Cambridge University Press, Cambridge},\ \bibinfo {year}
  {1995})\ Chap.\ \bibinfo {chapter} {9.6, 7.3}\BibitemShut {NoStop}%
\bibitem [{\citenamefont {Moffatt}(1978)}]{MoffattBOOK78}%
  \BibitemOpen
  \bibfield  {author} {\bibinfo {author} {\bibfnamefont {H.~K.}\ \bibnamefont
  {Moffatt}},\ }\href@noop {} {\emph {\bibinfo {title} {Magnetic Fild
  Generation in an Electrically Conducting Fluid}}}\ (\bibinfo  {publisher}
  {Cambridge University Press, Cambridge},\ \bibinfo {year} {1978})\BibitemShut
  {NoStop}%
\bibitem [{\citenamefont {Priest}\ and\ \citenamefont
  {Forbes}(2002)}]{PriestAAR02}%
  \BibitemOpen
  \bibfield  {author} {\bibinfo {author} {\bibfnamefont {E.~R.}\ \bibnamefont
  {Priest}}\ and\ \bibinfo {author} {\bibfnamefont {T.~G.}\ \bibnamefont
  {Forbes}},\ }\href@noop {} {\bibfield  {journal} {\bibinfo  {journal} {The
  Astron. Astrophys. Rev.}\ }\textbf {\bibinfo {volume} {10}},\ \bibinfo
  {pages} {313} (\bibinfo {year} {2002})}\BibitemShut {NoStop}%
\bibitem [{\citenamefont {Lin}\ \emph {et~al.}(1998)\citenamefont {Lin},
  \citenamefont {Hahm}, \citenamefont {Lee}, \citenamefont {Tang},\ and\
  \citenamefont {White}}]{LinScience98}%
  \BibitemOpen
  \bibfield  {author} {\bibinfo {author} {\bibfnamefont {Z.}~\bibnamefont
  {Lin}}, \bibinfo {author} {\bibfnamefont {T.~S.}\ \bibnamefont {Hahm}},
  \bibinfo {author} {\bibfnamefont {W.~W.}\ \bibnamefont {Lee}}, \bibinfo
  {author} {\bibfnamefont {W.~M.}\ \bibnamefont {Tang}}, \ and\ \bibinfo
  {author} {\bibfnamefont {R.~B.}\ \bibnamefont {White}},\ }\href@noop {}
  {\bibfield  {journal} {\bibinfo  {journal} {Science}\ }\textbf {\bibinfo
  {volume} {281}},\ \bibinfo {pages} {1835} (\bibinfo {year}
  {1998})}\BibitemShut {NoStop}%
\bibitem [{\citenamefont {Doyle}\ \emph {et~al.}(2007)\citenamefont {Doyle},
  \citenamefont {Houlberg}, \citenamefont {Kamada}, \citenamefont {Mukhovatov},
  \citenamefont {Osborne},\ and\ \citenamefont {et. al.}}]{DoyleNF07}%
  \BibitemOpen
  \bibfield  {author} {\bibinfo {author} {\bibfnamefont {E.~J.}\ \bibnamefont
  {Doyle}}, \bibinfo {author} {\bibfnamefont {W.~A.}\ \bibnamefont {Houlberg}},
  \bibinfo {author} {\bibfnamefont {Y.}~\bibnamefont {Kamada}}, \bibinfo
  {author} {\bibfnamefont {V.}~\bibnamefont {Mukhovatov}}, \bibinfo {author}
  {\bibfnamefont {T.~H.}\ \bibnamefont {Osborne}}, \ and\ \bibinfo {author}
  {\bibnamefont {et. al.}},\ }\href@noop {} {\bibfield  {journal} {\bibinfo
  {journal} {Nucl. Fusion}\ }\textbf {\bibinfo {volume} {47}},\ \bibinfo
  {pages} {S18} (\bibinfo {year} {2007})}\BibitemShut {NoStop}%
\bibitem [{\citenamefont {Bondeson}\ and\ \citenamefont
  {Ward}(1994)}]{BondesonPRL94}%
  \BibitemOpen
  \bibfield  {author} {\bibinfo {author} {\bibfnamefont {A.}~\bibnamefont
  {Bondeson}}\ and\ \bibinfo {author} {\bibfnamefont {D.~J.}\ \bibnamefont
  {Ward}},\ }\href@noop {} {\bibfield  {journal} {\bibinfo  {journal} {Phys.
  Rev. Lett.}\ }\textbf {\bibinfo {volume} {72}},\ \bibinfo {pages} {2709}
  (\bibinfo {year} {1994})}\BibitemShut {NoStop}%
\bibitem [{\citenamefont {Betti}\ and\ \citenamefont
  {Freidberg}(1994)}]{BettiPRL95}%
  \BibitemOpen
  \bibfield  {author} {\bibinfo {author} {\bibfnamefont {R.}~\bibnamefont
  {Betti}}\ and\ \bibinfo {author} {\bibfnamefont {J.~P.}\ \bibnamefont
  {Freidberg}},\ }\href@noop {} {\bibfield  {journal} {\bibinfo  {journal}
  {Phys. Rev. Lett.}\ }\textbf {\bibinfo {volume} {72}},\ \bibinfo {pages}
  {2709} (\bibinfo {year} {1994})}\BibitemShut {NoStop}%
\bibitem [{\citenamefont {Biglary}\ \emph {et~al.}(1990)\citenamefont
  {Biglary}, \citenamefont {Diamond},\ and\ \citenamefont
  {Terry}}]{BiglaryPoFB90}%
  \BibitemOpen
  \bibfield  {author} {\bibinfo {author} {\bibfnamefont {H.}~\bibnamefont
  {Biglary}}, \bibinfo {author} {\bibfnamefont {P.~H.}\ \bibnamefont
  {Diamond}}, \ and\ \bibinfo {author} {\bibfnamefont {P.~W.}\ \bibnamefont
  {Terry}},\ }\href@noop {} {\bibfield  {journal} {\bibinfo  {journal} {Phys.
  Fluids B}\ }\textbf {\bibinfo {volume} {2}},\ \bibinfo {pages} {1} (\bibinfo
  {year} {1990})}\BibitemShut {NoStop}%
\bibitem [{\citenamefont {Connor}\ and\ \citenamefont
  {Wilson}(1994)}]{ConnorPPCF94}%
  \BibitemOpen
  \bibfield  {author} {\bibinfo {author} {\bibfnamefont {J.~W.}\ \bibnamefont
  {Connor}}\ and\ \bibinfo {author} {\bibfnamefont {H.~R.}\ \bibnamefont
  {Wilson}},\ }\href@noop {} {\bibfield  {journal} {\bibinfo  {journal} {Plasma
  Phys. Contrl. Fusion}\ }\textbf {\bibinfo {volume} {36}},\ \bibinfo {pages}
  {719} (\bibinfo {year} {1994})}\BibitemShut {NoStop}%
\bibitem [{\citenamefont {Terry}(2000)}]{TerryRMP00}%
  \BibitemOpen
  \bibfield  {author} {\bibinfo {author} {\bibfnamefont {P.~W.}\ \bibnamefont
  {Terry}},\ }\href@noop {} {\bibfield  {journal} {\bibinfo  {journal} {Rev.
  Mod. Phys.}\ }\textbf {\bibinfo {volume} {72}},\ \bibinfo {pages} {109}
  (\bibinfo {year} {2000})}\BibitemShut {NoStop}%
\bibitem [{\citenamefont {Wang}(2006{\natexlab{a}})}]{WangPRL06a}%
  \BibitemOpen
  \bibfield  {author} {\bibinfo {author} {\bibfnamefont {S.}~\bibnamefont
  {Wang}},\ }\href@noop {} {\bibfield  {journal} {\bibinfo  {journal} {Phys.
  Rev. Lett.}\ }\textbf {\bibinfo {volume} {97}},\ \bibinfo {pages} {085002;
  129902} (\bibinfo {year} {2006}{\natexlab{a}})}\BibitemShut {NoStop}%
\bibitem [{\citenamefont {Peeters}\ \emph {et~al.}(2011)\citenamefont
  {Peeters}, \citenamefont {Angioni}, \citenamefont {Bortolon}, \citenamefont
  {Camenen}, \citenamefont {Casson}, \citenamefont {Duval}, \citenamefont
  {Fiederspiel}, \citenamefont {Hornsby}, \citenamefont {Idomura},
  \citenamefont {Hein}, \citenamefont {Kluy}, \citenamefont {Mantica},
  \citenamefont {Parra}, \citenamefont {Snodin}, \citenamefont {Szepesi}, ,
  \citenamefont {Strintzi}, \citenamefont {Tala}, \citenamefont {Tardini},
  \citenamefont {de~Vries},\ and\ \citenamefont {Weiland}}]{PeetersNF11}%
  \BibitemOpen
  \bibfield  {author} {\bibinfo {author} {\bibfnamefont {A.~G.}\ \bibnamefont
  {Peeters}}, \bibinfo {author} {\bibfnamefont {C.}~\bibnamefont {Angioni}},
  \bibinfo {author} {\bibfnamefont {A.}~\bibnamefont {Bortolon}}, \bibinfo
  {author} {\bibfnamefont {Y.}~\bibnamefont {Camenen}}, \bibinfo {author}
  {\bibfnamefont {F.~J.}\ \bibnamefont {Casson}}, \bibinfo {author}
  {\bibfnamefont {B.}~\bibnamefont {Duval}}, \bibinfo {author} {\bibfnamefont
  {L.}~\bibnamefont {Fiederspiel}}, \bibinfo {author} {\bibfnamefont {W.~A.}\
  \bibnamefont {Hornsby}}, \bibinfo {author} {\bibfnamefont {Y.}~\bibnamefont
  {Idomura}}, \bibinfo {author} {\bibfnamefont {T.}~\bibnamefont {Hein}},
  \bibinfo {author} {\bibfnamefont {N.}~\bibnamefont {Kluy}}, \bibinfo {author}
  {\bibfnamefont {P.}~\bibnamefont {Mantica}}, \bibinfo {author} {\bibfnamefont
  {F.~I.}\ \bibnamefont {Parra}}, \bibinfo {author} {\bibfnamefont {A.~P.}\
  \bibnamefont {Snodin}}, \bibinfo {author} {\bibfnamefont {G.}~\bibnamefont
  {Szepesi}}, , \bibinfo {author} {\bibfnamefont {D.}~\bibnamefont {Strintzi}},
  \bibinfo {author} {\bibfnamefont {T.}~\bibnamefont {Tala}}, \bibinfo {author}
  {\bibfnamefont {G.}~\bibnamefont {Tardini}}, \bibinfo {author} {\bibfnamefont
  {P.}~\bibnamefont {de~Vries}}, \ and\ \bibinfo {author} {\bibfnamefont
  {J.}~\bibnamefont {Weiland}},\ }\href@noop {} {\bibfield  {journal} {\bibinfo
   {journal} {Nucl. Fusion}\ }\textbf {\bibinfo {volume} {51}},\ \bibinfo
  {pages} {094027} (\bibinfo {year} {2011})}\BibitemShut {NoStop}%
\bibitem [{\citenamefont {Angioni}\ \emph {et~al.}(2012)\citenamefont
  {Angioni}, \citenamefont {Camenen}, \citenamefont {Casson}, \citenamefont
  {Fable}, \citenamefont {McDermott}, \citenamefont {Peeters},\ and\
  \citenamefont {Rice}}]{AngioniNF12}%
  \BibitemOpen
  \bibfield  {author} {\bibinfo {author} {\bibfnamefont {C.}~\bibnamefont
  {Angioni}}, \bibinfo {author} {\bibfnamefont {Y.}~\bibnamefont {Camenen}},
  \bibinfo {author} {\bibfnamefont {F.~J.}\ \bibnamefont {Casson}}, \bibinfo
  {author} {\bibfnamefont {F.}~\bibnamefont {Fable}}, \bibinfo {author}
  {\bibfnamefont {R.~M.}\ \bibnamefont {McDermott}}, \bibinfo {author}
  {\bibfnamefont {A.~G.}\ \bibnamefont {Peeters}}, \ and\ \bibinfo {author}
  {\bibfnamefont {J.~E.}\ \bibnamefont {Rice}},\ }\href@noop {} {\bibfield
  {journal} {\bibinfo  {journal} {Nucl. Fusion}\ }\textbf {\bibinfo {volume}
  {52}},\ \bibinfo {pages} {114003} (\bibinfo {year} {2012})}\BibitemShut
  {NoStop}%
\bibitem [{\citenamefont {Lee}\ \emph {et~al.}(2003)\citenamefont {Lee},
  \citenamefont {Rice}, \citenamefont {Marmar}, \citenamefont {Greenwald},
  \citenamefont {Hutchinson},\ and\ \citenamefont {Snipes}}]{LeePRL03}%
  \BibitemOpen
  \bibfield  {author} {\bibinfo {author} {\bibfnamefont {W.~D.}\ \bibnamefont
  {Lee}}, \bibinfo {author} {\bibfnamefont {J.~E.}\ \bibnamefont {Rice}},
  \bibinfo {author} {\bibfnamefont {E.~S.}\ \bibnamefont {Marmar}}, \bibinfo
  {author} {\bibfnamefont {M.~J.}\ \bibnamefont {Greenwald}}, \bibinfo {author}
  {\bibfnamefont {I.~H.}\ \bibnamefont {Hutchinson}}, \ and\ \bibinfo {author}
  {\bibfnamefont {J.~A.}\ \bibnamefont {Snipes}},\ }\href@noop {} {\bibfield
  {journal} {\bibinfo  {journal} {Phys. Rev. Lett.}\ }\textbf {\bibinfo
  {volume} {91}},\ \bibinfo {pages} {205003} (\bibinfo {year}
  {2003})}\BibitemShut {NoStop}%
\bibitem [{\citenamefont {Rice}\ \emph {et~al.}(2007)\citenamefont {Rice},
  \citenamefont {Ince-Cushman}, \citenamefont {deGrassie}, \citenamefont
  {Eriksson}, \citenamefont {Sakamoto}, \citenamefont {Scarabosio},
  \citenamefont {Bortolon}, \citenamefont {Burrell}, \citenamefont {Duval},
  \citenamefont {Fenzi-Bonizec}, \citenamefont {Greenwald}, \citenamefont
  {Groebner}, \citenamefont {Hoang}, \citenamefont {Koide}, \citenamefont
  {Marmar}, \citenamefont {Pochelon},\ and\ \citenamefont
  {Podpaly}}]{RiceNF07}%
  \BibitemOpen
  \bibfield  {author} {\bibinfo {author} {\bibfnamefont {J.~E.}\ \bibnamefont
  {Rice}}, \bibinfo {author} {\bibfnamefont {A.}~\bibnamefont {Ince-Cushman}},
  \bibinfo {author} {\bibfnamefont {J.~S.}\ \bibnamefont {deGrassie}}, \bibinfo
  {author} {\bibfnamefont {L.~G.}\ \bibnamefont {Eriksson}}, \bibinfo {author}
  {\bibfnamefont {Y.}~\bibnamefont {Sakamoto}}, \bibinfo {author}
  {\bibfnamefont {A.}~\bibnamefont {Scarabosio}}, \bibinfo {author}
  {\bibfnamefont {A.}~\bibnamefont {Bortolon}}, \bibinfo {author}
  {\bibfnamefont {K.~H.}\ \bibnamefont {Burrell}}, \bibinfo {author}
  {\bibfnamefont {B.~P.}\ \bibnamefont {Duval}}, \bibinfo {author}
  {\bibfnamefont {C.}~\bibnamefont {Fenzi-Bonizec}}, \bibinfo {author}
  {\bibfnamefont {M.~J.}\ \bibnamefont {Greenwald}}, \bibinfo {author}
  {\bibfnamefont {R.~J.}\ \bibnamefont {Groebner}}, \bibinfo {author}
  {\bibfnamefont {G.~T.}\ \bibnamefont {Hoang}}, \bibinfo {author}
  {\bibfnamefont {Y.}~\bibnamefont {Koide}}, \bibinfo {author} {\bibfnamefont
  {E.~S.}\ \bibnamefont {Marmar}}, \bibinfo {author} {\bibfnamefont
  {A.}~\bibnamefont {Pochelon}}, \ and\ \bibinfo {author} {\bibfnamefont
  {Y.}~\bibnamefont {Podpaly}},\ }\href@noop {} {\bibfield  {journal} {\bibinfo
   {journal} {Nucl. Fusion}\ }\textbf {\bibinfo {volume} {47}},\ \bibinfo
  {pages} {1618} (\bibinfo {year} {2007})}\BibitemShut {NoStop}%
\bibitem [{\citenamefont {Rice}\ \emph {et~al.}(2008)\citenamefont {Rice},
  \citenamefont {Ince-Cushman}, \citenamefont {Reinke}, \citenamefont
  {Podpaly}, \citenamefont {Greenwald}, \citenamefont {LaBombard},\ and\
  \citenamefont {Marmar}}]{RicePPCF08}%
  \BibitemOpen
  \bibfield  {author} {\bibinfo {author} {\bibfnamefont {J.~E.}\ \bibnamefont
  {Rice}}, \bibinfo {author} {\bibfnamefont {A.}~\bibnamefont {Ince-Cushman}},
  \bibinfo {author} {\bibfnamefont {M.~L.}\ \bibnamefont {Reinke}}, \bibinfo
  {author} {\bibfnamefont {Y.}~\bibnamefont {Podpaly}}, \bibinfo {author}
  {\bibfnamefont {M.~J.}\ \bibnamefont {Greenwald}}, \bibinfo {author}
  {\bibfnamefont {B.}~\bibnamefont {LaBombard}}, \ and\ \bibinfo {author}
  {\bibfnamefont {E.~S.}\ \bibnamefont {Marmar}},\ }\href@noop {} {\bibfield
  {journal} {\bibinfo  {journal} {Plasma Phys. Control. Fusion}\ }\textbf
  {\bibinfo {volume} {50}},\ \bibinfo {pages} {124042} (\bibinfo {year}
  {2008})}\BibitemShut {NoStop}%
\bibitem [{\citenamefont {Solomon}\ \emph {et~al.}(2007)\citenamefont
  {Solomon}, \citenamefont {Burrell}, \citenamefont {deGrassie}, \citenamefont
  {Budny}, \citenamefont {Groebner}, \citenamefont {Kinsey}, \citenamefont
  {Kramer}, \citenamefont {Luce}, \citenamefont {Makowski}, \citenamefont
  {Mikkelsen}, \citenamefont {Nazikian}, \citenamefont {Petty}, \citenamefont
  {Politzer}, \citenamefont {Scott}, \citenamefont {Van~Zeeland},\ and\
  \citenamefont {Zarnstorff}}]{SolomonPPCF07}%
  \BibitemOpen
  \bibfield  {author} {\bibinfo {author} {\bibfnamefont {W.~M.}\ \bibnamefont
  {Solomon}}, \bibinfo {author} {\bibfnamefont {K.~H.}\ \bibnamefont
  {Burrell}}, \bibinfo {author} {\bibfnamefont {J.~S.}\ \bibnamefont
  {deGrassie}}, \bibinfo {author} {\bibfnamefont {R.}~\bibnamefont {Budny}},
  \bibinfo {author} {\bibfnamefont {R.~J.}\ \bibnamefont {Groebner}}, \bibinfo
  {author} {\bibfnamefont {J.~E.}\ \bibnamefont {Kinsey}}, \bibinfo {author}
  {\bibfnamefont {G.~J.}\ \bibnamefont {Kramer}}, \bibinfo {author}
  {\bibfnamefont {T.~C.}\ \bibnamefont {Luce}}, \bibinfo {author}
  {\bibfnamefont {M.~A.}\ \bibnamefont {Makowski}}, \bibinfo {author}
  {\bibfnamefont {D.}~\bibnamefont {Mikkelsen}}, \bibinfo {author}
  {\bibfnamefont {R.}~\bibnamefont {Nazikian}}, \bibinfo {author}
  {\bibfnamefont {C.~C.}\ \bibnamefont {Petty}}, \bibinfo {author}
  {\bibfnamefont {P.~A.}\ \bibnamefont {Politzer}}, \bibinfo {author}
  {\bibfnamefont {S.~D.}\ \bibnamefont {Scott}}, \bibinfo {author}
  {\bibfnamefont {M.~A.}\ \bibnamefont {Van~Zeeland}}, \ and\ \bibinfo {author}
  {\bibfnamefont {M.~C.}\ \bibnamefont {Zarnstorff}},\ }\href@noop {}
  {\bibfield  {journal} {\bibinfo  {journal} {Plasma Phys. Contrl. Fusion}\
  }\textbf {\bibinfo {volume} {49}},\ \bibinfo {pages} {B313} (\bibinfo {year}
  {2007})}\BibitemShut {NoStop}%
\bibitem [{\citenamefont {Solomon}\ \emph {et~al.}(2009)\citenamefont
  {Solomon}, \citenamefont {Burrell}, \citenamefont {Garofalo}, \citenamefont
  {Cole}, \citenamefont {Budny}, \citenamefont {deGrassie}, \citenamefont
  {Heidbrink}, \citenamefont {Jackson}, \citenamefont {Lanctot}, \citenamefont
  {Nazikian}, \citenamefont {Reimerdes}, \citenamefont {Strait}, \citenamefont
  {Van~Zeeland},\ and\ \citenamefont {the DIII-D Rotation Physics
  Task~Force}}]{SolomonNF09}%
  \BibitemOpen
  \bibfield  {author} {\bibinfo {author} {\bibfnamefont {W.~M.}\ \bibnamefont
  {Solomon}}, \bibinfo {author} {\bibfnamefont {K.~H.}\ \bibnamefont
  {Burrell}}, \bibinfo {author} {\bibfnamefont {A.~M.}\ \bibnamefont
  {Garofalo}}, \bibinfo {author} {\bibfnamefont {A.~J.}\ \bibnamefont {Cole}},
  \bibinfo {author} {\bibfnamefont {R.~V.}\ \bibnamefont {Budny}}, \bibinfo
  {author} {\bibfnamefont {J.~S.}\ \bibnamefont {deGrassie}}, \bibinfo {author}
  {\bibfnamefont {W.~W.}\ \bibnamefont {Heidbrink}}, \bibinfo {author}
  {\bibfnamefont {G.~L.}\ \bibnamefont {Jackson}}, \bibinfo {author}
  {\bibfnamefont {M.~J.}\ \bibnamefont {Lanctot}}, \bibinfo {author}
  {\bibfnamefont {R.}~\bibnamefont {Nazikian}}, \bibinfo {author}
  {\bibfnamefont {H.}~\bibnamefont {Reimerdes}}, \bibinfo {author}
  {\bibfnamefont {E.~J.}\ \bibnamefont {Strait}}, \bibinfo {author}
  {\bibfnamefont {M.~A.}\ \bibnamefont {Van~Zeeland}}, \ and\ \bibinfo {author}
  {\bibnamefont {the DIII-D Rotation Physics Task~Force}},\ }\href@noop {}
  {\bibfield  {journal} {\bibinfo  {journal} {Nucl. Fusion}\ }\textbf {\bibinfo
  {volume} {49}},\ \bibinfo {pages} {085005} (\bibinfo {year}
  {2009})}\BibitemShut {NoStop}%
\bibitem [{\citenamefont {Peeters}\ \emph {et~al.}(2007)\citenamefont
  {Peeters}, \citenamefont {Angioni},\ and\ \citenamefont
  {Strintzi}}]{PeetersPRL07}%
  \BibitemOpen
  \bibfield  {author} {\bibinfo {author} {\bibfnamefont {A.~G.}\ \bibnamefont
  {Peeters}}, \bibinfo {author} {\bibfnamefont {C.}~\bibnamefont {Angioni}}, \
  and\ \bibinfo {author} {\bibfnamefont {D.}~\bibnamefont {Strintzi}},\
  }\href@noop {} {\bibfield  {journal} {\bibinfo  {journal} {Phys. Rev. Lett.}\
  }\textbf {\bibinfo {volume} {98}},\ \bibinfo {pages} {265003} (\bibinfo
  {year} {2007})}\BibitemShut {NoStop}%
\bibitem [{\citenamefont {Hahm}\ \emph {et~al.}(2007)\citenamefont {Hahm},
  \citenamefont {Diamond}, \citenamefont {Gurcan},\ and\ \citenamefont
  {Rewoldt}}]{HahmPoP07}%
  \BibitemOpen
  \bibfield  {author} {\bibinfo {author} {\bibfnamefont {T.~S.}\ \bibnamefont
  {Hahm}}, \bibinfo {author} {\bibfnamefont {P.~H.}\ \bibnamefont {Diamond}},
  \bibinfo {author} {\bibfnamefont {O.~D.}\ \bibnamefont {Gurcan}}, \ and\
  \bibinfo {author} {\bibfnamefont {G.}~\bibnamefont {Rewoldt}},\ }\href@noop
  {} {\bibfield  {journal} {\bibinfo  {journal} {Phys. Plasmas}\ }\textbf
  {\bibinfo {volume} {14}},\ \bibinfo {pages} {072302} (\bibinfo {year}
  {2007})}\BibitemShut {NoStop}%
\bibitem [{\citenamefont {Diamond}\ \emph {et~al.}(2008)\citenamefont
  {Diamond}, \citenamefont {McDevitt}, \citenamefont {Gurcan}, \citenamefont
  {Hahm},\ and\ \citenamefont {Naulin}}]{DiamondPoP08}%
  \BibitemOpen
  \bibfield  {author} {\bibinfo {author} {\bibfnamefont {P.~H.}\ \bibnamefont
  {Diamond}}, \bibinfo {author} {\bibfnamefont {C.~J.}\ \bibnamefont
  {McDevitt}}, \bibinfo {author} {\bibfnamefont {O.~D.}\ \bibnamefont
  {Gurcan}}, \bibinfo {author} {\bibfnamefont {T.~S.}\ \bibnamefont {Hahm}}, \
  and\ \bibinfo {author} {\bibfnamefont {V.}~\bibnamefont {Naulin}},\
  }\href@noop {} {\bibfield  {journal} {\bibinfo  {journal} {Phys. Plasmas}\
  }\textbf {\bibinfo {volume} {15}},\ \bibinfo {pages} {012303} (\bibinfo
  {year} {2008})}\BibitemShut {NoStop}%
\bibitem [{\citenamefont {Garbet}\ \emph {et~al.}(2013)\citenamefont {Garbet},
  \citenamefont {Esteve}, \citenamefont {Sarazin}, \citenamefont {Abiteboul},
  \citenamefont {Bourdelle}, \citenamefont {Dif-Pradalier}, \citenamefont
  {Ghendrih}, \citenamefont {Grandgirard}, \citenamefont {Latu},\ and\
  \citenamefont {Smolyakov}}]{GarbetPoP13}%
  \BibitemOpen
  \bibfield  {author} {\bibinfo {author} {\bibfnamefont {X.}~\bibnamefont
  {Garbet}}, \bibinfo {author} {\bibfnamefont {D.}~\bibnamefont {Esteve}},
  \bibinfo {author} {\bibfnamefont {Y.}~\bibnamefont {Sarazin}}, \bibinfo
  {author} {\bibfnamefont {J.}~\bibnamefont {Abiteboul}}, \bibinfo {author}
  {\bibfnamefont {C.}~\bibnamefont {Bourdelle}}, \bibinfo {author}
  {\bibfnamefont {G.}~\bibnamefont {Dif-Pradalier}}, \bibinfo {author}
  {\bibfnamefont {P.}~\bibnamefont {Ghendrih}}, \bibinfo {author}
  {\bibfnamefont {V.}~\bibnamefont {Grandgirard}}, \bibinfo {author}
  {\bibfnamefont {G.}~\bibnamefont {Latu}}, \ and\ \bibinfo {author}
  {\bibfnamefont {A.}~\bibnamefont {Smolyakov}},\ }\href@noop {} {\bibfield
  {journal} {\bibinfo  {journal} {Phys. Plasmas}\ }\textbf {\bibinfo {volume}
  {20}},\ \bibinfo {pages} {072502} (\bibinfo {year} {2013})}\BibitemShut
  {NoStop}%
\bibitem [{\citenamefont {Galeev}\ and\ \citenamefont
  {Sagdeev}(1968)}]{GaleevJETP68}%
  \BibitemOpen
  \bibfield  {author} {\bibinfo {author} {\bibfnamefont {A.~A.}\ \bibnamefont
  {Galeev}}\ and\ \bibinfo {author} {\bibfnamefont {R.~Z.}\ \bibnamefont
  {Sagdeev}},\ }\href@noop {} {\bibfield  {journal} {\bibinfo  {journal}
  {Soviet Physics JETP}\ }\textbf {\bibinfo {volume} {26}},\ \bibinfo {pages}
  {233} (\bibinfo {year} {1968})}\BibitemShut {NoStop}%
\bibitem [{\citenamefont {Bickerton}\ \emph {et~al.}(1971)\citenamefont
  {Bickerton}, \citenamefont {Connor},\ and\ \citenamefont
  {Taylor}}]{BickertonNPS71}%
  \BibitemOpen
  \bibfield  {author} {\bibinfo {author} {\bibfnamefont {R.~J.}\ \bibnamefont
  {Bickerton}}, \bibinfo {author} {\bibfnamefont {J.~W.}\ \bibnamefont
  {Connor}}, \ and\ \bibinfo {author} {\bibfnamefont {J.~B.}\ \bibnamefont
  {Taylor}},\ }\href@noop {} {\bibfield  {journal} {\bibinfo  {journal} {Nature
  Phys. Science}\ }\textbf {\bibinfo {volume} {229}},\ \bibinfo {pages} {110}
  (\bibinfo {year} {1971})}\BibitemShut {NoStop}%
\bibitem [{\citenamefont {Hazeltine}\ and\ \citenamefont
  {Meiss}(1992)}]{HazeltineBOOK92}%
  \BibitemOpen
  \bibfield  {author} {\bibinfo {author} {\bibfnamefont {R.~D.}\ \bibnamefont
  {Hazeltine}}\ and\ \bibinfo {author} {\bibfnamefont {J.~D.}\ \bibnamefont
  {Meiss}},\ }\href@noop {} {\emph {\bibinfo {title} {Plasma Confinement}}}\
  (\bibinfo  {publisher} {Addison-Wesley Publishing Company},\ \bibinfo {year}
  {1992})\ Chap.\ \bibinfo {chapter} {6, 8}, pp.\ \bibinfo {pages} {195,
  368}\BibitemShut {NoStop}%
\bibitem [{\citenamefont {Callen}\ \emph {et~al.}(2009)\citenamefont {Callen},
  \citenamefont {Cole},\ and\ \citenamefont {Hegna}}]{CallenNF09}%
  \BibitemOpen
  \bibfield  {author} {\bibinfo {author} {\bibfnamefont {J.~D.}\ \bibnamefont
  {Callen}}, \bibinfo {author} {\bibfnamefont {A.~J.}\ \bibnamefont {Cole}}, \
  and\ \bibinfo {author} {\bibfnamefont {C.~C.}\ \bibnamefont {Hegna}},\
  }\href@noop {} {\bibfield  {journal} {\bibinfo  {journal} {Nucl. Fusion}\
  }\textbf {\bibinfo {volume} {49}},\ \bibinfo {pages} {085021} (\bibinfo
  {year} {2009})}\BibitemShut {NoStop}%
\bibitem [{\citenamefont {Hirshman}\ and\ \citenamefont
  {Sigmar}(1981)}]{HirshmanNF81}%
  \BibitemOpen
  \bibfield  {author} {\bibinfo {author} {\bibfnamefont {S.~P.}\ \bibnamefont
  {Hirshman}}\ and\ \bibinfo {author} {\bibfnamefont {D.~J.}\ \bibnamefont
  {Sigmar}},\ }\href@noop {} {\bibfield  {journal} {\bibinfo  {journal} {Nucl.
  Fusion}\ }\textbf {\bibinfo {volume} {21}},\ \bibinfo {pages} {1079}
  (\bibinfo {year} {1981})}\BibitemShut {NoStop}%
\bibitem [{\citenamefont {Zarnstorff}\ \emph {et~al.}(1988)\citenamefont
  {Zarnstorff}, \citenamefont {Bell}, \citenamefont {Bitter}, \citenamefont
  {Goldston}, \citenamefont {Grek}, \citenamefont {Hawryluk}, \citenamefont
  {Hill}, \citenamefont {Johnson}, \citenamefont {Mccune}, \citenamefont
  {Park}, \citenamefont {Ramsey}, \citenamefont {Taylor},\ and\ \citenamefont
  {Wieland}}]{ZarnstorffPRL88}%
  \BibitemOpen
  \bibfield  {author} {\bibinfo {author} {\bibfnamefont {M.~C.}\ \bibnamefont
  {Zarnstorff}}, \bibinfo {author} {\bibfnamefont {M.~G.}\ \bibnamefont
  {Bell}}, \bibinfo {author} {\bibfnamefont {M.}~\bibnamefont {Bitter}},
  \bibinfo {author} {\bibfnamefont {R.~J.}\ \bibnamefont {Goldston}}, \bibinfo
  {author} {\bibfnamefont {B.}~\bibnamefont {Grek}}, \bibinfo {author}
  {\bibfnamefont {R.~J.}\ \bibnamefont {Hawryluk}}, \bibinfo {author}
  {\bibfnamefont {K.}~\bibnamefont {Hill}}, \bibinfo {author} {\bibfnamefont
  {D.}~\bibnamefont {Johnson}}, \bibinfo {author} {\bibfnamefont
  {D.}~\bibnamefont {Mccune}}, \bibinfo {author} {\bibfnamefont
  {H.}~\bibnamefont {Park}}, \bibinfo {author} {\bibfnamefont {A.}~\bibnamefont
  {Ramsey}}, \bibinfo {author} {\bibfnamefont {G.}~\bibnamefont {Taylor}}, \
  and\ \bibinfo {author} {\bibfnamefont {R.}~\bibnamefont {Wieland}},\
  }\href@noop {} {\bibfield  {journal} {\bibinfo  {journal} {Phys. Rev. Lett.}\
  }\textbf {\bibinfo {volume} {60}},\ \bibinfo {pages} {1306} (\bibinfo {year}
  {1988})}\BibitemShut {NoStop}%
\bibitem [{\citenamefont {Hwang}\ \emph {et~al.}(1996)\citenamefont {Hwang},
  \citenamefont {Forest},\ and\ \citenamefont {Ono}}]{HwangPRL96}%
  \BibitemOpen
  \bibfield  {author} {\bibinfo {author} {\bibfnamefont {Y.~S.}\ \bibnamefont
  {Hwang}}, \bibinfo {author} {\bibfnamefont {C.~B.}\ \bibnamefont {Forest}}, \
  and\ \bibinfo {author} {\bibfnamefont {M.}~\bibnamefont {Ono}},\ }\href@noop
  {} {\bibfield  {journal} {\bibinfo  {journal} {Phys. Rev. Lett.}\ }\textbf
  {\bibinfo {volume} {77}},\ \bibinfo {pages} {3811} (\bibinfo {year}
  {1996})}\BibitemShut {NoStop}%
\bibitem [{\citenamefont {Thomas}\ \emph {et~al.}(2004)\citenamefont {Thomas},
  \citenamefont {Leonard}, \citenamefont {Lao},\ and\ \citenamefont
  {Osborne}}]{ThomasPRL04}%
  \BibitemOpen
  \bibfield  {author} {\bibinfo {author} {\bibfnamefont {D.~M.}\ \bibnamefont
  {Thomas}}, \bibinfo {author} {\bibfnamefont {A.~W.}\ \bibnamefont {Leonard}},
  \bibinfo {author} {\bibfnamefont {L.~L.}\ \bibnamefont {Lao}}, \ and\
  \bibinfo {author} {\bibfnamefont {T.~H.}\ \bibnamefont {Osborne}},\
  }\href@noop {} {\bibfield  {journal} {\bibinfo  {journal} {Phys. Rev. Lett.}\
  }\textbf {\bibinfo {volume} {93}},\ \bibinfo {pages} {065003} (\bibinfo
  {year} {2004})}\BibitemShut {NoStop}%
\bibitem [{\citenamefont {Wade}\ \emph {et~al.}(2004)\citenamefont {Wade},
  \citenamefont {Murakami},\ and\ \citenamefont {Polizter}}]{WadePRL04}%
  \BibitemOpen
  \bibfield  {author} {\bibinfo {author} {\bibfnamefont {M.~R.}\ \bibnamefont
  {Wade}}, \bibinfo {author} {\bibfnamefont {M.}~\bibnamefont {Murakami}}, \
  and\ \bibinfo {author} {\bibfnamefont {P.~A.}\ \bibnamefont {Polizter}},\
  }\href@noop {} {\bibfield  {journal} {\bibinfo  {journal} {Phys. Rev. Lett.}\
  }\textbf {\bibinfo {volume} {92}},\ \bibinfo {pages} {235005} (\bibinfo
  {year} {2004})}\BibitemShut {NoStop}%
\bibitem [{\citenamefont {Kagan}\ and\ \citenamefont
  {Catto}(2010)}]{KaganPRL10}%
  \BibitemOpen
  \bibfield  {author} {\bibinfo {author} {\bibfnamefont {G.}~\bibnamefont
  {Kagan}}\ and\ \bibinfo {author} {\bibfnamefont {P.~J.}\ \bibnamefont
  {Catto}},\ }\href@noop {} {\bibfield  {journal} {\bibinfo  {journal} {Phys.
  Rev. Lett.}\ }\textbf {\bibinfo {volume} {105}},\ \bibinfo {pages} {045002}
  (\bibinfo {year} {2010})}\BibitemShut {NoStop}%
\bibitem [{\citenamefont {McDevitt}\ \emph {et~al.}(2013)\citenamefont
  {McDevitt}, \citenamefont {Tang},\ and\ \citenamefont {Guo}}]{McDevittPRL13}%
  \BibitemOpen
  \bibfield  {author} {\bibinfo {author} {\bibfnamefont {C.~J.}\ \bibnamefont
  {McDevitt}}, \bibinfo {author} {\bibfnamefont {X.}~\bibnamefont {Tang}}, \
  and\ \bibinfo {author} {\bibfnamefont {Z.}~\bibnamefont {Guo}},\ }\href@noop
  {} {\bibfield  {journal} {\bibinfo  {journal} {Phys. Rev. Lett.}\ }\textbf
  {\bibinfo {volume} {111}},\ \bibinfo {pages} {205002} (\bibinfo {year}
  {2013})}\BibitemShut {NoStop}%
\bibitem [{\citenamefont {Kikuchi}\ and\ \citenamefont
  {Azumi}(2012)}]{KikuchiRMP12}%
  \BibitemOpen
  \bibfield  {author} {\bibinfo {author} {\bibfnamefont {M.}~\bibnamefont
  {Kikuchi}}\ and\ \bibinfo {author} {\bibfnamefont {M.}~\bibnamefont
  {Azumi}},\ }\href@noop {} {\bibfield  {journal} {\bibinfo  {journal} {Rev.
  Mod. Phys.}\ }\textbf {\bibinfo {volume} {84}},\ \bibinfo {pages} {1807}
  (\bibinfo {year} {2012})}\BibitemShut {NoStop}%
\bibitem [{\citenamefont {Wang}(2012{\natexlab{a}})}]{WangPPCF12}%
  \BibitemOpen
  \bibfield  {author} {\bibinfo {author} {\bibfnamefont {S.}~\bibnamefont
  {Wang}},\ }\href@noop {} {\bibfield  {journal} {\bibinfo  {journal} {Plasma
  Phys. Control. Fusion}\ }\textbf {\bibinfo {volume} {54}},\ \bibinfo {pages}
  {015003} (\bibinfo {year} {2012}{\natexlab{a}})}\BibitemShut {NoStop}%
\end{thebibliography}

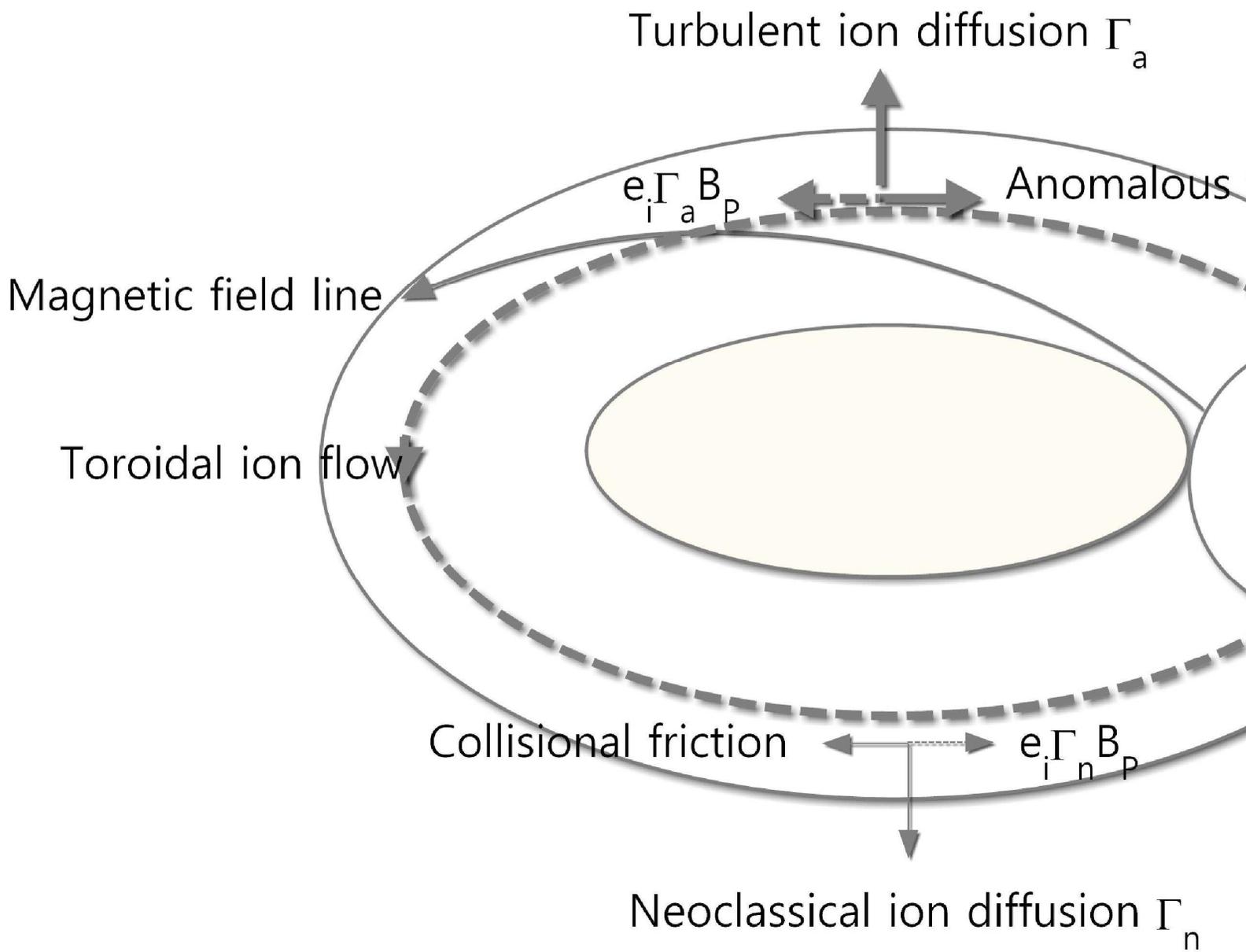
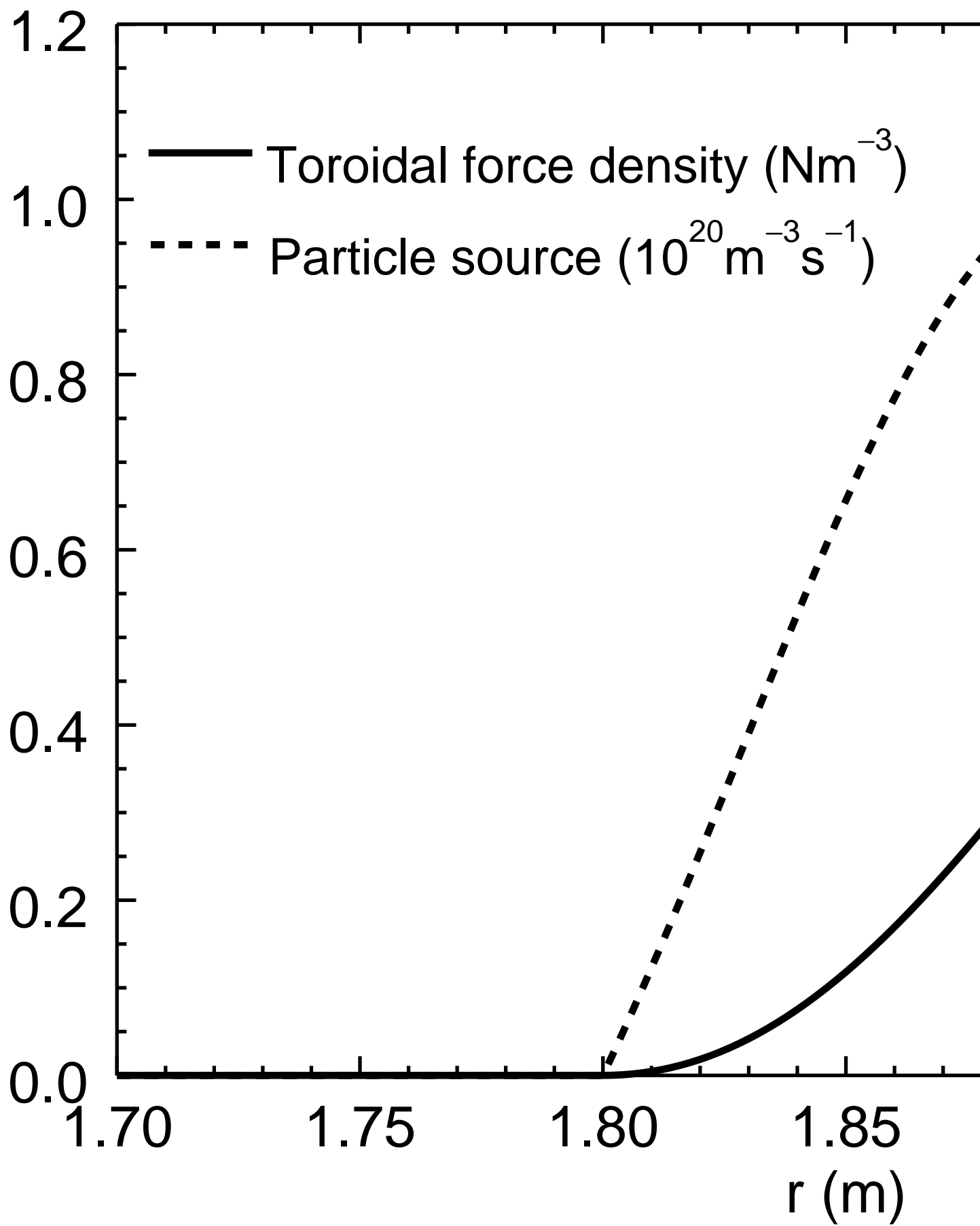
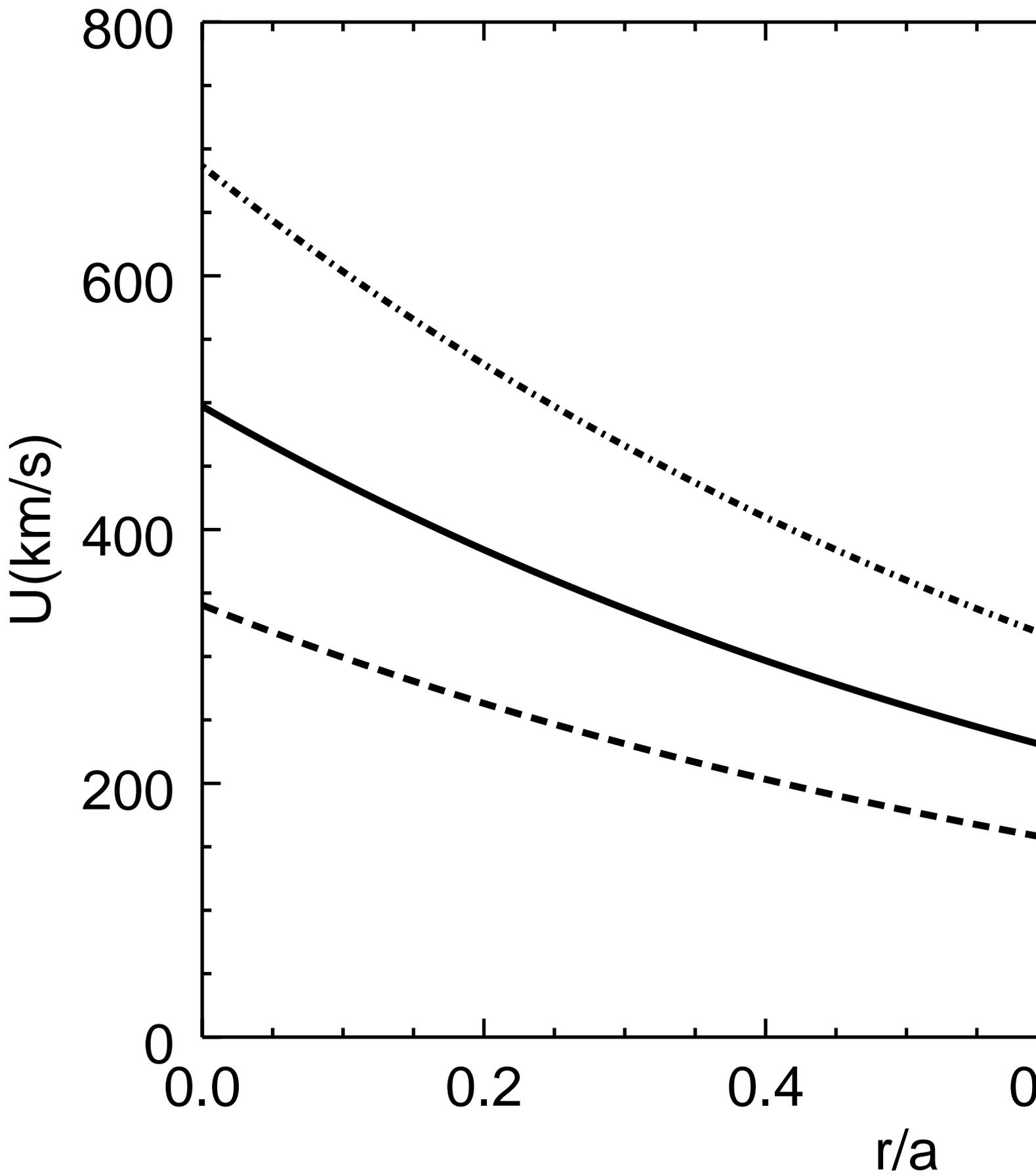
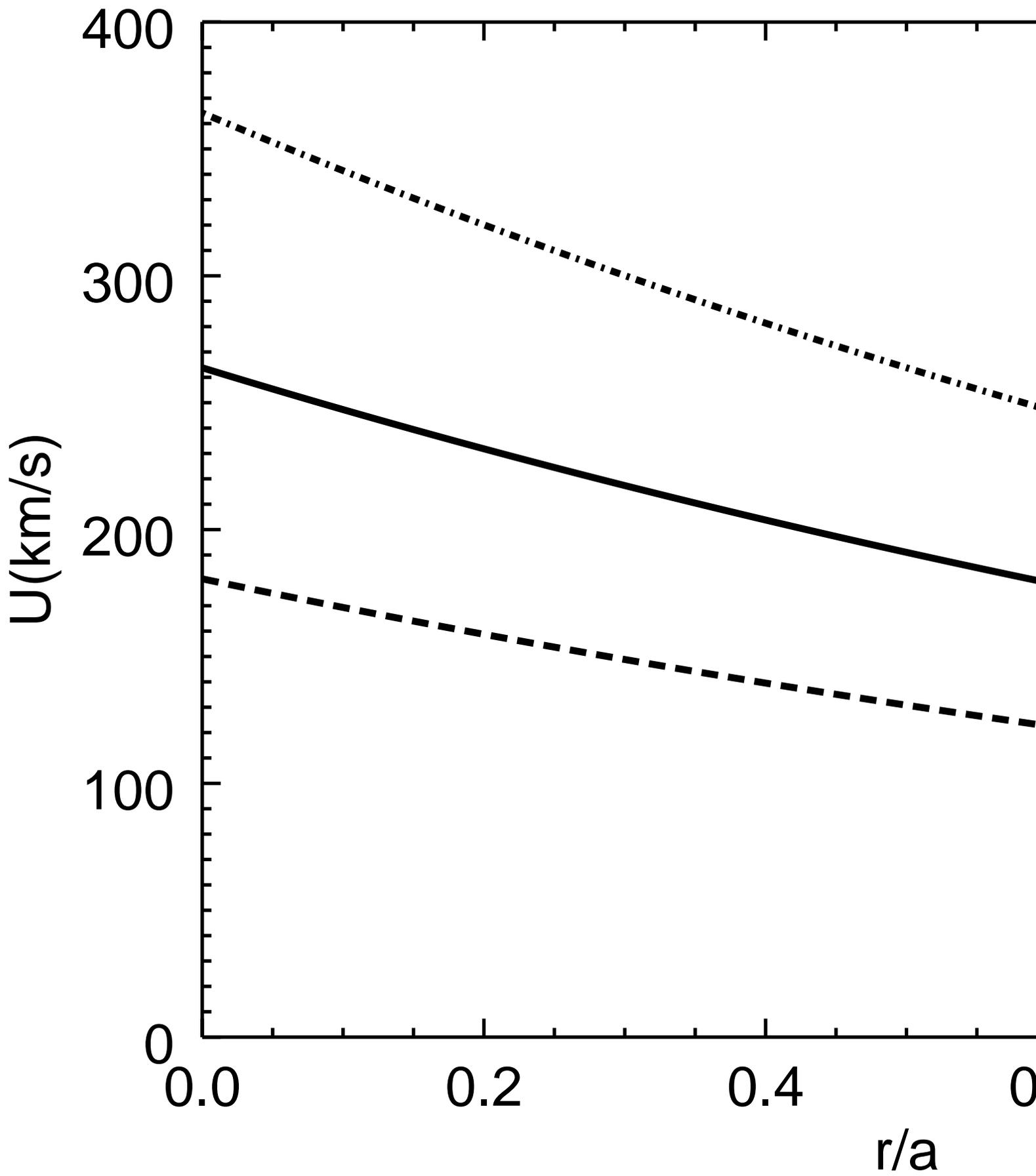
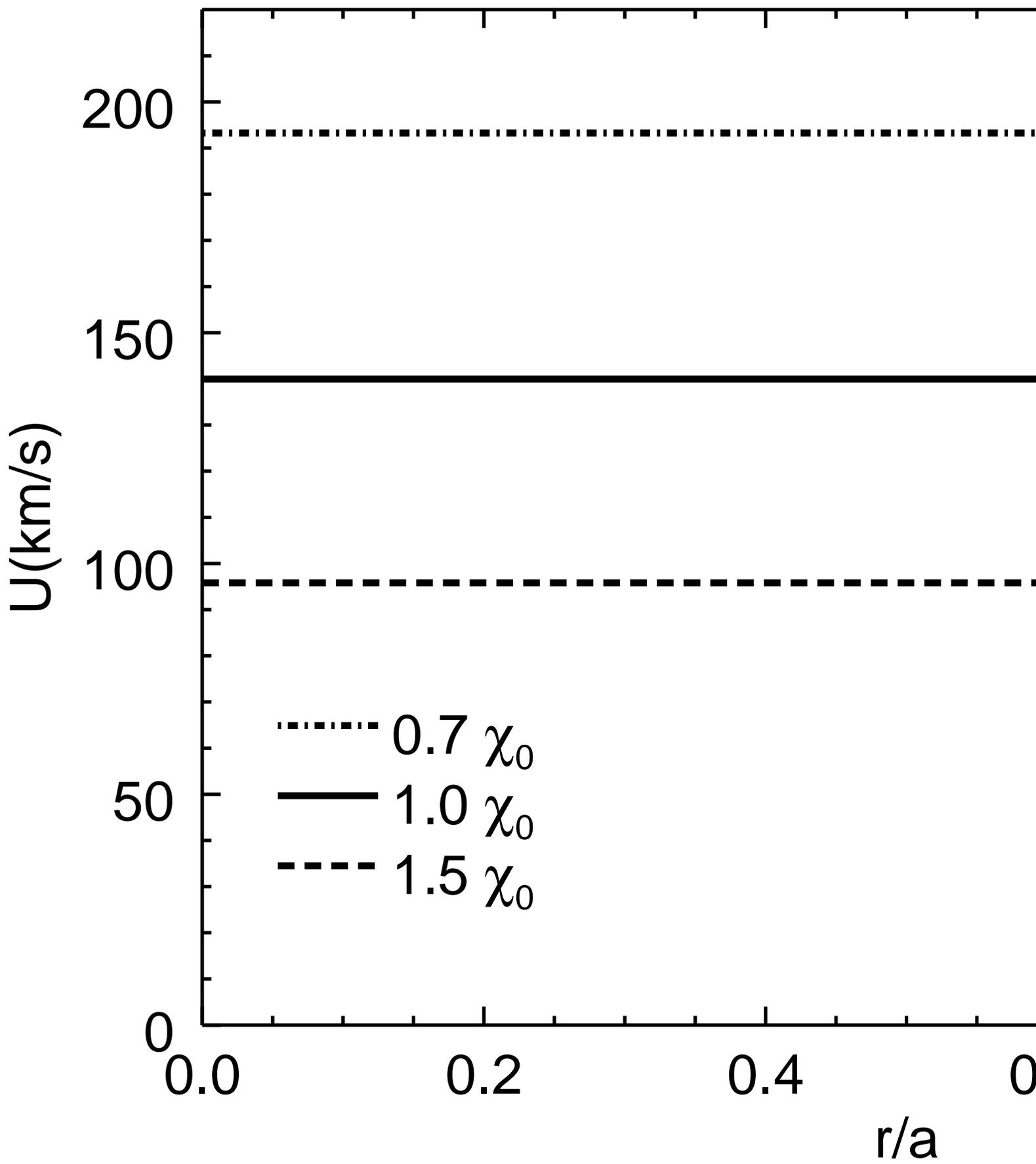
\end{document}